\documentclass[journal]{IEEEtran}
\ifCLASSINFOpdf
  \usepackage[pdftex]{graphicx}
  \DeclareGraphicsExtensions{.pdf,.jpeg,.png}
\else
  \usepackage[dvips]{graphicx}
  \DeclareGraphicsExtensions{.eps}
\fi
\usepackage{epstopdf}
\usepackage{tikz}
\usepackage{url}
\usepackage[cmex10]{amsmath}
\usepackage{cleveref}
\usepackage{mathtools} 
\usepackage{color}

\usepackage{cite}

\ifCLASSOPTIONcompsoc
  \usepackage[caption=false,font=normalsize,labelfont=sf,textfont=sf,justification=centering]{subfig}
\else
  \usepackage[caption=false,font=footnotesize,justification=centering]{subfig}
\fi
\usepackage{stfloats}
\renewcommand*{\Re}{\operatorname{Re}} 
\renewcommand*{\Im}{\operatorname{Im}}

\usepackage{algorithm}
\usepackage{algpseudocode}
\usepackage{amsfonts}
\usepackage{widetext}
\usepackage{float}
\usepackage{multirow}

\algnewcommand\algorithmicinput{\textbf{Input:}}
\algnewcommand\Input{\item[\algorithmicinput]}
\algnewcommand\algorithmicoutput{\textbf{Output:}}
\algnewcommand\Output{\item[\algorithmicoutput]}

\begin{document}
%
\title{Augmented Generator Sub-transient Model\\ Using Dynamic Phasor Measurements}
%
%
%

\author{Pablo Marchi, Francisco Messina, 
Leonardo Rey Vega, 
Cecilia G. Galarza\thanks{The work of P. Marchi  was supported by a CONICET Ph.D. grant. This work was partially funded by the UREE 4 FONARSEC project: ``Development of Synchrophasor Measurements for Smart Electrical Grids''. The FONARSEC is funded by the Ministry of Science, Technology and Innovaton of Argentina.}
\thanks{P. Marchi, F. Messina, L. Rey Vega and C. G. Galarza are with the CSC-CONICET, and the School of Engineering - Universidad de Buenos Aires, Argentina (e-mail: pmarchi@csc.conicet.gov.ar, fmessina@fi.uba.ar, lrey@fi.uba.ar, cgalar@fi.uba.ar).}}%
\maketitle

\begin{abstract}
In this article, we present a new model for a synchronous generator based on phasor measurement units (PMUs) data. The proposed sub-transient model allows to estimate the dynamic state variables as well as to calibrate model parameters. The motivation for this new model is to use more efficiently the PMU measurements which are becoming widely available in power grids. The concept of phasor derivative is applied, which not only includes the signal phase derivative but also its amplitude derivative. Applying known non-linear estimation techniques, we study the merits of this new model. In particular, we test robustness by considering a generator with different mechanical power controls.
\end{abstract}


%
\IEEEpeerreviewmaketitle

\section{Introduction}
%
%
%
%

\IEEEPARstart{D}{ynamic}  models form the basis for power system transient stability
simulations. Simulation accuracy is driven, in part, by the accuracy
of the individual models used to represent actual equipments installed
in the field. Models are developed during the baseline testing in close coordination between the generator
owner and manufacturer of the components. However, modeling errors exist
in the dynamic studies used for planning and operating the bulk power
system. These errors are introduced through component replacements,
aging, measurement error, etc., that are not captured in these preliminary 
models. Some historical disturbances can partly be attributed to model
inaccuracy. Post-mortem analyses using the ideal model from the planning stage
have shown gross differences from actual performance \cite{Nerc2015}. Nowadays, the North American
Electric Reliability Corporation (NERC) Reliability Standards MOD-026-1,
MOD-027-1, MOD-032-1 and MOD-033-1 \cite{NERC} seek to ensure that dynamic models remain within pre-defined limits so that they accurately represent the equipment installed in the field.

Traditionally, a short-circuit test on unloaded synchronous generator units offered the standard measure for transient parameters. However, due to its limitations on providing q-axis transient and sub-transient constants, several alternative tests, such
as enhanced sudden short circuit test, stator decrement test and standstill frequency response test, have been proposed for obtaining a better representation of the dynamic model \cite{Syngen2003}. Nevertheless, in practice, the implementation of these offline methods is inconvenient due to the high cost incurred when performing the disconnection of the generators. Recently, online methods have been proposed in order to assess the dynamic behavior, and to reduce the uncertainties, when the generator is working under stressful conditions. These techniques were designed to harness the measurements from Digital Fault Recorders (DFRs) installed at the point of connection \cite{Tsai2017}.

The online methods can be separated in two groups depending on the type of data processing. The first one uses a frame-based processing approach. Some pioneering works on this matter are \cite{Sanchez1988,Burth1999}. In \cite{Sanchez1988}, the identification of synchronous generator reactances and time constants of the excitation system is achieved using a trajectory sensitivity method. In \cite{Burth1999}, nonlinear least squares estimation is applied to obtain a subset of the model parameters. With the inclusion of Phasor Measurement Units (PMUs), these techniques have evolved to include these new measurements. Among other advantages, these electronic devices record the electromechanical dynamics of the generating units with good precision and high reporting rates, which can reach up to 120 frames per cycle. That was how the second group of processing techniques arose. They are based on sample by sample Bayesian filtering, such as: the Extended Kalman Filter (EKF)\cite{Huang2009}, Unscented Kalman Filter (UKF)\cite{AGHAMOLKI201545}, Ensemble Kalman Filter (EnKF)\cite{Fan2015}, and Particle Filter (PF) \cite{Zhou2015}.

The main objective of this paper is not to discuss which estimation technique is more adequate, but instead to analyze the model to be considered. Concretely, a sub-transient model is adopted and a new way of defining the transition function of the model is explained. It is worth mentioning that the sub-transient model was chosen with the criteria of including as many physical effects as possible. Thus, the performance of the estimates is improved by reducing the number of model uncertainties.
\\
\\
The main contributions of this work are:

\begin{itemize}
\item The presentation of a sub-transient generator model with the possibility
of including the Automatic Voltage Regulator (AVR), the Power System Stabilizer (PSS) and the Turbine Governor (TG) control loops.
\item A novel model that include not only voltage and current phasors, but also their time derivatives as well as the frequency and Rate Of Change Of Frequency (ROCOF).
\end{itemize}
\vspace{0,8cm}
The paper is organized as follows. Following a description
of the model parameter estimation problem in Section II, we discuss the formulation of
the sub-transient generator model in Section III.
Section IV presents different scenarios to show the validity
of the proposed method. Section V presents the conclusions and possible future work directions.

\section{Problem Formulation}

\subsection{Problem statement}
\begin{figure}[t]
\centering
\includegraphics[width = 1\columnwidth]{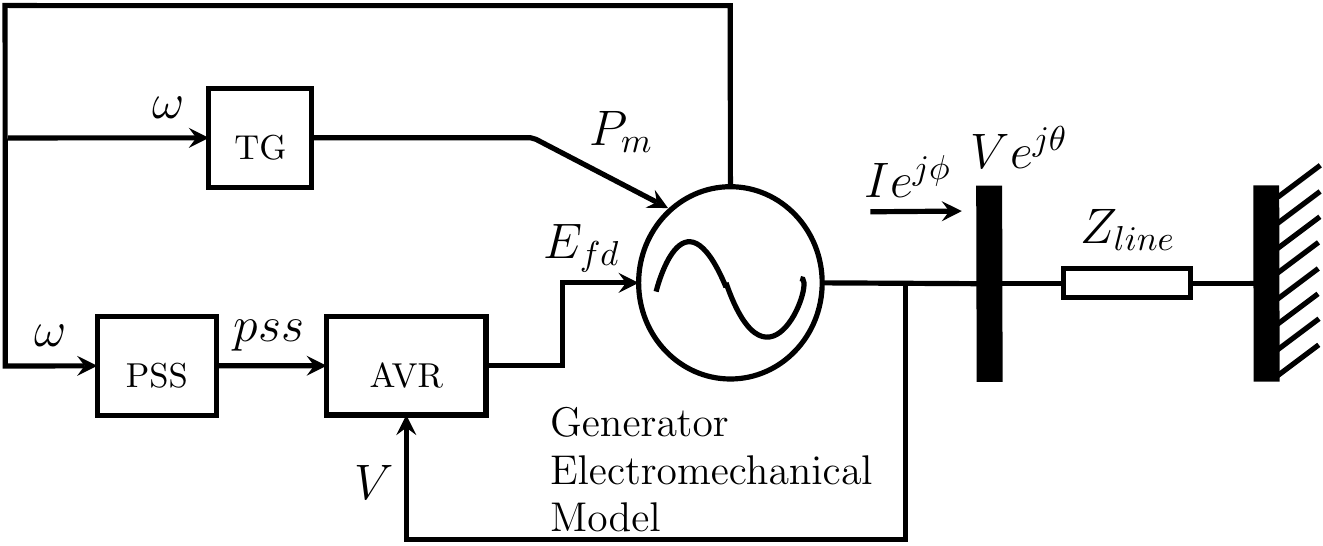}
\caption{Generator with controllers connected to an infinite bus.}
\label{fig:gen_controllers}
\end{figure}

The goal is to identify the model parameters as well as to dynamically estimate the internal states of a generator unit. To achieve this, measurements from a PMU at the point of connection will be used as inputs to the filtering algorithm. This type of technique is widely used and is known as \textit{event playback} \cite{Huang2013}. To introduce the subject, Fig. \ref{fig:gen_controllers} shows a general structure of a power plant. The system consists of a synchronous generator, a TG, an AVR and a PSS. 

\subsection{Dynamic State Estimation}

To estimate the dynamics of a generator unit, the general form of a state-space model for nonlinear systems is considered:
\begin{equation}
\begin{aligned}
\mathbf{x}_{k} &= f_s(\mathbf{x}_{k-1},\mathbf{u}_{k-1})+\mathbf{w}_{k},\,\,\,\,\mathbf{x}_{k}\in\mathbb{R}^n,\\
\mathbf{z}_{k} &= h_s(\mathbf{x}_{k},\mathbf{u}_{k-1})+\mathbf{v}_{k}.
\end{aligned}
\label{eq:nonlinearsystem}
\end{equation}
where $f_s$ is the state transition function which models the generators dynamics, $\mathbf{x}$ is 
the state vector, $\mathbf{z}$ is the measurement vector, $\mathbf{u}$ is the input vector, $h_s$ is the function which relates the measurements with the state vector, and $\mathbf{w}$ and $\mathbf{v}$ are noise vectors introduced to account for modeling errors.

\subsection{Review of Unscented Kalman Filter}

Given its simplicity, its reduced computational cost, and its good performance for non-linear systems, we  have adopted the UKF estimation algorithm. In addition, as it was shown \mbox{in \cite{Zhou2015}}, the UKF is a feasible real time solution. 

Unlike the well-known EKF, the UKF gets certain amount of extra terms from the Taylor series of $f_s$ and $h_s$. Besides, it has not the necessity of computing the Jacobian of these functions. This algorithm defines the Unscented transformation to approximate the mean and covariance of the state vector. For this purpose, the concept of \textit{sigma points} is introduced. 
These sigma points are propagated through the nonlinear functions 
and then the mean and the covariance for $\mathbf{x}$ ($\hat{\mathbf{x}}_k$, ${{P_k}}$) and $\mathbf{z}$ ($\hat{\mathbf{z}}_k$, ${{H_k}}$) are 
approximated using a weighted sample mean and covariance of the posterior sigma 
points. Basically, the procedure can be divided in two steps: prediction 
and correction. Equations \eqref{eq:UKF0} to \eqref{eq:UKF3} summarize the filtering algorithm.\\
\\
\textit{Select three positive scalars $\gamma,\,\, \beta,\,\, \kappa$ and define the following constants}:
\begin{subequations}
\begin{align}
\,\,\, \lambda &=\gamma^2(n+\kappa)-n\,,\\
w^m_0 &= \frac{\lambda}{n+\lambda},\,\,\, {w}^c_0 = \frac{\lambda}{n+\lambda}+(1-\gamma^2+\beta), \\
w^m_h &= w^c_h = \frac{1}{2(n+\lambda)}\,\,\,,h=1,\cdots,2n.
\end{align} \label{eq:UKF0}
\end{subequations}

\noindent
\textit{Prediction step}:\\

Predicted (\textit{a priori}) state estimate,
\begin{subequations}
\begin{align}
\hat{\mathbf{x}}_{k|k-1}^{i} & =f_s\left(\hat{\mathbf{x}}_{k-1|k-1}^{i},\mathbf{u}_{k-1}\right),\\
\hat{\mathbf{x}}_{k-1|k-1}^{0} & =\hat{\mathbf{x}}_{k-1|k-1},\\
\hat{\mathbf{x}}_{k-1|k-1}^{j} & =\hat{\mathbf{x}}_{k-1|k-1}+\left(\sqrt{\left(n+\lambda\right)P_{k-1|k-1}}\right)_{j}\,,\\
\hat{\mathbf{x}}_{k-1|k-1}^{j+n} & =\hat{\mathbf{x}}_{k-1|k-1}\,-\left(\sqrt{\left(n+\lambda\right)P_{k-1|k-1}}\right)_{j}\,,\\
&\hspace{4,03cm}j=1,\cdots,n,\nonumber\\
\hat{\mathbf{x}}_{k|k-1} &=\sum_{i=0}^{2n} w^m_{i}\,\hat{\mathbf{x}}_{k|k-1}^{i}.
\end{align} \label{eq:UKF1}
\end{subequations}

Predicted (\textit{a priori}) state covariance,
\begin{subequations}
\begin{align}
P_{k|k-1} &= \tilde{P}_{k|k-1} + Q_{k},\\
\tilde{P}_{k|k-1} &=\sum_{i=0}^{2n} w^c_{i}\,\left(\hat{\mathbf{x}}_{k|k-1}^{i}-\hat{\mathbf{x}}_{k|k-1}\right)\left(\hat{\mathbf{x}}_{k|k-1}^{i}-\hat{\mathbf{x}}_{k|k-1}\right)^{T}.
\end{align} \label{eq:UKF2}
\end{subequations}
\noindent
\textit{Correction step}:
\begin{subequations}
\begin{align}
\hat{\mathbf{z}}_{k|k-1}^{i}&=h_s\left(\hat{\mathbf{x}}_{k|k-1}^{i},\mathbf{u}_{k-1}\right),\\
\hat{\mathbf{z}}_{k|k-1}&=\sum_{i=0}^{2n} w^m_{i}\,\hat{\mathbf{z}}_{k|k-1}^{i},\\
\tilde{\mathbf{y}}_{k}&=\mathbf{z}_{k}-\hat{\mathbf{z}}_{k|k-1},\\
H_{k}&=\sum_{i=0}^{2n} w^c_{i}\,\tilde{\mathbf{y}}_{k}\,\tilde{\mathbf{y}}_{k}^{T},\\
K_{k}&=P_{k|k-1}H_{k}^{T}\left(H_{k}P_{k|k-1}H_{k}^{T}+R_{k}\right)^{-1},\\
\hat{\mathbf{x}}_{k|k}&=\hat{\mathbf{x}}_{k|k-1}+K_{k}\tilde{\mathbf{y}}_{k},\\
P_{k|k}&=\left(I-K_{k}H_{k}\right)P_{k|k-1}.
\end{align} \label{eq:UKF3}
\end{subequations}
Here, the estimate of the state vector at time $k$ is computed using the measurements at time $l$ and is denoted as $\hat{\mathbf{x}}_{k|l}$, $\hat{\mathbf{x}}_{l|m}^{i}$ and $\hat{\mathbf{z}}_{l|m}^{i}$  are the sigma points of the state vector and the measurements respectively $\forall\, i=0,\cdots,2n$, $R_k$ is the measurement noise covariance matrix at time $k$, $Q_k$ the process noise covariance matrix at time $k$, and the $\left( \, \right)_j$ operator takes the $j$-th row of the matrix.   For an in-depth discussion, please refer to \cite{AGHAMOLKI201545}.

\section{Dynamic Generator Modeling}

\subsection{Conventional Sub-transient Generator Model} \label{sec:sub_gen}

The transient model is the simplest model that allows to add control loops into the mechanical power and field voltage. Because of that, it is widely used \cite{Zhou2015,Ghahremani2016,Zhao2017}. However, the sub-transient model is more complete since it incorporates to the transient model a new set of time constants ($T''_d$, $T''_q$) that define faster electromagnetic changes. The model was introduced for calibration purposes in \cite{Huang2013}, but only the rotor equations were considered. In this paper, all the effects modeled in the Power System Toolbox (PST) \cite{PST} are taken into account. This analysis lays the basis for the model that will be introduced in Section \ref{sec:improvement}. Consider the following set of equations expressed in the per-unit (p.u.) system:
\\
\\
\hspace*{0.03cm} Electromechanical equations:
\begin{subequations}
\begin{align}
\frac{d\delta}{dt} & = \omega_s\,\left(\omega-\omega_{0}\right),\\
\frac{d\omega}{dt} & =\frac{\omega_{0}}{2\,H}\,\left[P_{m}-T_{e}-D\,\left(\omega-\omega_{0}\right)\right]. \label{eq:sub_tra1}
\end{align} 
\end{subequations}
Subtransitent rotor equations: 
\begin{subequations}
\begin{align}
\frac{dE'_{d}}{dt} & =\frac{1}{T'_{q}}\left[-E_{d}-k_2\left(E'_{d}-\Psi_{q}\right)-k_1\,I_{q}\right],\\
\frac{dE'_{q}}{dt} & =\frac{1}{T'_{d}}\left[E_{fd}-S(E'_q)-k_3\left(E'_{q}-\Psi_{d}\right)-k_4\,I_{d}\right],\\
\frac{d\Psi_{d}}{dt} & =\frac{1}{T''_{d}}\left[-\Psi_{d}+E'_{q}-\left(x'_{d}-x_{ls}\right)\,I_{d}\right],\\
\frac{d\Psi_{q}}{dt} & =\frac{1}{T''_{q}}\left[-\Psi_{q}+E'_{d}-\left(x'_{q}-x_{ls}\right)\,I_{q}\right].
\end{align} \label{eq:sub_tra2}
\end{subequations}
AVR equations:
\begin{subequations}
\begin{align}
\frac{dE_{fd}}{dt} & = \frac{1}{T_{A}}\,\left[-E_{fd} + K_A \left(pss + V_{REF} - V_{TR} \right)\right],\\
\frac{dV_{TR}}{dt} & =\frac{1}{T_{R}}\left(V-V_{TR}\right).
\end{align} \label{eq:sub_tra3}
\end{subequations}
TG equation:
\begin{align}
\frac{dP_{m}}{dt} & = \frac{1}{T_{ef}}\,\left[-P_{m} + \left(1-\omega\right) \frac{1}{r} + P_{m,0} \right]. 
\label{eq:sub_tra4}
\end{align} 

To complete the system, additional equations are given in \mbox{Appendix A}. A full description of the notation is given in \mbox{Appendix B}. The model presented above can be found in \cite{sauer2006power}. As in \cite{Zhao2017} the AVR was modeled as a proportional-integral control, and the transducer effect has been taken into account. Unlike \cite{AGHAMOLKI201545} and \cite{Zhao2017}, the TG equation only considers a simple pole defined by an effective time constant $T_{ef}$. This consideration will be discussed later. To simplify the exposition, the PSS was modeled as a constant. Alternatively, its output could be included in (\ref{eq:sub_tra3}a).

To select the parameters of the generator to be estimated we refer to the sensitivity analysis carried out in \cite{Tsai2012}, and corroborated by \cite{Huang2013}. These \textit{key parameters} will be those whose deviations produce greater changes in the delivered active and reactive power. Thus, the parameters to estimate are the inertia constant and the exciter
gain, defining the parameter vector to calibrate $\mathbf{x}_{cal}=[H,\, K_A]$. Therefore, the state vector
is defined as:
\begin{equation}
\mathbf{x}=\left[\delta\,\,\,\omega\,\,\,E'_{d}\,\,\,E'_{q}\,\,\,\Psi_{d}\,\,\,\Psi_{q}\,\,\,E_{fd}\,\,\,V_{TR}\,\,\,P_m\,\,\,\mathbf{x}_{cal} \right]^{T},
\end{equation}

\noindent
and $\dot{\mathbf{x}}{}_{cal}=\mathbf{0}$ is used to complete the specification of $f_s$.

Using
the same criteria as in {\cite{Zhou2015}},
the measurement vector is composed of the real and imaginary parts of the voltage phasor:
\begin{equation}
\mathbf{z}=\left[V_{re}\,V_{im}\right]^{T}.
\end{equation}
Under this criteria, $\mathbf{z}_k$ will contain an associated noise whose distribution corresponds to the distribution of the error in the voltage phasor measurement. To relate these measurements to the state vector, the electrical interface for this type of models is used (see Fig. \ref{fig:subtransient_gen}). Using the following definitions:
\begin{align*}
\Psi''_{q} & =\frac{x_{ls}-x_{q}''}{x_{q}'-x_{ls}}E'_{d}-\frac{x_{q}'-x_{q}''}{x_{q}'-x_{ls}}\Psi_{q},\\
\Psi''_{d}&=\left(\frac{x_{d}''-x_{ls}}{x_{d}'-x_{ls}}E'_{q}+\frac{x_{d}'-x_{d}''}{x_{d}'-x_{ls}}\Psi_{d}\right),
\end{align*}
and assuming that $x''_{q}\approx x''_{d}$, the current flow through the branch can be expressed as: 
\begin{figure}[]
\includegraphics[width = 1\columnwidth]{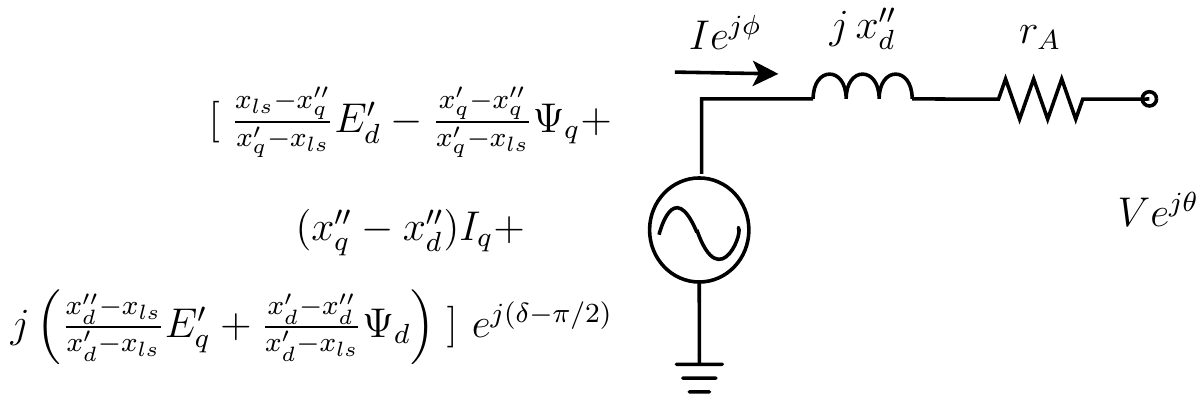}
\caption{Synchronous machine sub-transient dynamic circuit model.}
\label{fig:subtransient_gen}
\end{figure}
\begin{equation}
Ie^{j \phi}=\frac{\left(\Psi''_{q}+j\Psi''_{d}\right)\,e^{j(\delta-\pi/2)}-Ve^{j \theta}}{r_A+jx''_d}.
\end{equation}
After some manipulations, the measurements equations are:
\begin{subequations}
\begin{align}
V_{re}&=\Psi''_{q}\,\sin\left(\delta\right)+\Psi''_{d}\,\cos\left(\delta\right)+I_{im}x''_{d}-I_{re}r_{A},\\
V_{im}&=-\Psi''_{q}\,\cos\left(\delta\right)+\Psi''_{d}\,\sin\left(\delta\right)-I_{re}x''_{d}-I_{im}r_{A}.
\end{align} \label{eq:sub_med}
\end{subequations}
The resulting $h_s$ can be obtained from (\ref{eq:sub_med}). At last, the control vector necessary to stabilize the system  will be composed by the following signals:
\begin{equation}
\mathbf{u}=\left[P_{e}\,I_{re}\,I_{im}\right]^{T}.
\end{equation}

\subsection{Augmented Sub-transient Generator Model} \label{sec:improvement}
The idea is to extend the conventional model to include the frequency and the ROCOF measurements given by the PMUs. The standard definition of this quantities can be found in \cite{ieeestd2011}. In this way, we would expect that this model would increase the observability of the closed loop system. But the state variables are hard to relate with the measured frequency and ROCOF. Note that the value of these measurements depend on the dynamics of all the generators and associated loads in the entire power system. As our main interest is to perform the calibration procedure in a decoupled way, no accurate model can be proposed. However, as all the generators have
internal impedances much smaller
than the equivalent impedance of the rest of the network, the approximation
given by \eqref{eq:sub_approx} can be made. From now on, the ROCOF is denoted as $\alpha$.
\begin{equation}
\begin{array}{cc}
f & \approx\omega,\,\,\,\alpha\approx\dot{\omega}.
\end{array} \label{eq:sub_approx}
\end{equation}
Accordingly, the new measurement vector is defined as:
\begin{equation}
\mathbf{z}=\left[V_{re}\,\,V_{im}\,\,f\,\,\alpha\right]^{T} .
\end{equation}
Now, the function $h_s$ is defined using \eqref{eq:sub_med} and \eqref{eq:sub_approx}. Note that the quality of this approximation can be controlled by the selection of $R_k$. Besides, it is concluded that the variable $\dot{\omega}=\frac{d\omega}{dt}$
should be added to the state vector. To achieve this, the transition equation \eqref{eq:sub_tra1}
 should be modified:
\begin{subequations}
\begin{align}
\frac{d\omega}{dt} & =\dot{\omega},\\
\frac{d\dot{\omega}}{dt} &=\frac{\omega_{0}}{2\,H}\,\left(\dot{P}{}_{m}-\dot{T}_{e}-D\,\dot{\omega}\right). \label{eq:sub_tra_fr}
\end{align}
\end{subequations}
where, from \eqref{eq:electric_torque}, we obtain:
\begin{equation}
\dot{T}_{e}=\dot{P}_{e}+r_{A}\,\left(2\,I_{d}\,\dot{I}_{d}+2\,I_{q}\,\dot{I}_{q}\right). \label{eq:te_dot}
\end{equation}
By definition, and after differentiation we get:
\begin{equation}
P_{e}=\Re\left\{ S\right\} =\Re\left\{ Ve^{j\theta}\,\,Ie^{-j\phi}\right\} =V\,I\,\cos(\theta-\phi),
\end{equation}
\begin{align}
\dot{P}_{e}&=\left(\dot{V}\,I+V\,\dot{I}\right)\cos\left(\theta-\phi\right) 
-V\,I\,\sin\left(\theta-\phi\right)\left(\dot{\theta}-\dot{\phi}\right). \label{eq:dpe}
\end{align}

Now, the term of the derivative of the electrical torque which contains the losses of the armature resistance is analyzed. Taking into account \eqref{eq:id} and \eqref{eq:iq}, we obtain:
\begin{align}
\dot{I}_{d} & =\left(\dot{I}_{re}+I_{im}\, \omega\right)\sin(\delta)+\left(I_{re} \,\omega-\dot{I}_{im}\right)\cos(\delta), \label{eq:id_dot}\\
\dot{I}_{q} & =\left(\dot{I}_{im}-I_{re}\, \omega\right)\sin(\delta)+\left(I_{im} \,\omega+\dot{I}_{re}\right)\cos(\delta). \label{eq:iq_dot}
\end{align}
From \eqref{eq:te_dot}, \eqref{eq:id_dot} and \eqref{eq:iq_dot}, and after some simplifications we get:
\begin{equation}
\dot{T_{e}}=\dot{P_{e}}+2\,r_{A}\,\left(I_{re}\dot{I}_{re}+I_{im}\dot{I}_{im}\right).
\end{equation}

From \eqref{eq:sub_tra_fr}, we see that this model works with $\dot{P}_m$, instead of $P_m$, so \eqref{eq:sub_tra4} is replaced by:
\begin{align}
\frac{d\dot{P}_{m}}{dt} & = \frac{1}{T_{ef}}\,\left[-\dot{P}{}_{m} - \dot{\omega} \frac{1}{r} \right]. \label{eq:sub_tra_fr2}
\end{align} 
If it is necessary to estimate $P_m$, the state vector  $\mathbf{x}$ should be augmented again to include this new variable. Finally, the new control and state vectors are defined as:
\begin{align}
\mathbf{x}=&\left[\delta\,\,\omega\,\,\dot{\omega}\,\,E'_{d}\,\,E'_{q}\,\,\Psi_{d}\,\,\Psi_{q}\,\,E_{fd}\,\,V_{TR}\,\,\dot{P}_m\,\,\mathbf{x}_{cal} \right]^{T},\\
\mathbf{u}=&\left[\overrightarrow{V}\,\overrightarrow{I}\,\dot{\overrightarrow{V}}\,\dot{\overrightarrow{I}}\right]^{T}.
\end{align}

This model has three advantages: 
\begin{itemize}
\item The model does not depend on the mechanical power value in steady state $P_{m,0}$. In the literature, this value, as well as others parameters from TG detailed models, are assumed to be known. Nevertheless, in practice, this is not always the case.
\item The calibration process
is more robust. This can be recognized by inspecting (\ref{eq:sub_tra1}) and \eqref{eq:sub_tra_fr}. In both equations the variable $H$ is involved. In the first equation $P_m$ and $T_e$ are comparable magnitudes while in the last one $\dot{P}_m$ and $\dot{T}_e$ are not. In fact, $\dot{P}_m$ is always much smaller than $\dot{T}_e$. Accordingly, it is expected that this improvement could handle more sophisticated models without knowledge of the structure of the TG or its parameters.
\item By including the phasor derivatives at the input, the model gives more details of the dynamics of the rotor. Indeed, for a judiciously chosen description of the measurement noise, the estimates depend more on the transition model than on 
the measurement one. As a consequence, the approximation given by (\ref{eq:sub_approx}) becomes less relevant.  
\end{itemize}

\subsection{Phase and magnitude derivatives}

The PMU or DFR should be capable of measuring all the variables involved in \eqref{eq:dpe}. Beyond the fact that some of these magnitudes are not defined by the aforementioned standards, it is well known that there are several algorithms for phasor estimation which estimate the phasor and its first and second derivatives \cite{Serna2007,Petri2014,Bertocco2015,Messina2017}. From the phasor derivatives, it is possible to compute the derivatives for amplitude and phase in \eqref{eq:dpe}. If the voltage phasor is a complex number defined as $\overrightarrow{V}=V_{re}+jV_{im}=V\,e^{j\theta}$, then:
\begin{align}
\dot{\overrightarrow{V}}\equiv\frac{d\overrightarrow{V}}{dt}=&\left(\frac{dV}{dt}+j\,V\,\frac{d\theta}{dt}\right)\,e^{j\theta},\\
\frac{\dot{\overrightarrow{V}}}{\overrightarrow{V}}\,=&\frac{\dot{V}}{{V}}\,\,+j\,\frac{d\theta}{dt}.
\end{align}

The same procedure is followed for the current phasor $\overrightarrow{I}=I_{re}+jI_{im}=I\,e^{j\phi}$, and the following expressions are obtained:
\begin{equation}
\begin{cases}
\frac{dV}{dt} & =V\,\Re\left\{ \frac{\dot{\overrightarrow{V}}}{\overrightarrow{V}}\right\} \\
\frac{d\theta}{dt} & =\Im\left\{ \frac{\dot{\overrightarrow{V}}}{\overrightarrow{V}}\right\} 
\end{cases},\,\,\,\,\,\,\begin{cases}
\frac{dI}{dt} & =I\,\Re\left\{\frac{\dot{\overrightarrow{I}}}{\overrightarrow{I}}\right\} \\
\frac{d\phi}{dt} & =\Im\left\{\frac{\dot{\overrightarrow{I}}}{\overrightarrow{I}}\right\} 
\end{cases}
\end{equation}

\section{Numerical results} \label{sec:results}

\begin{figure}[t]
\includegraphics[width = 0.95\columnwidth]{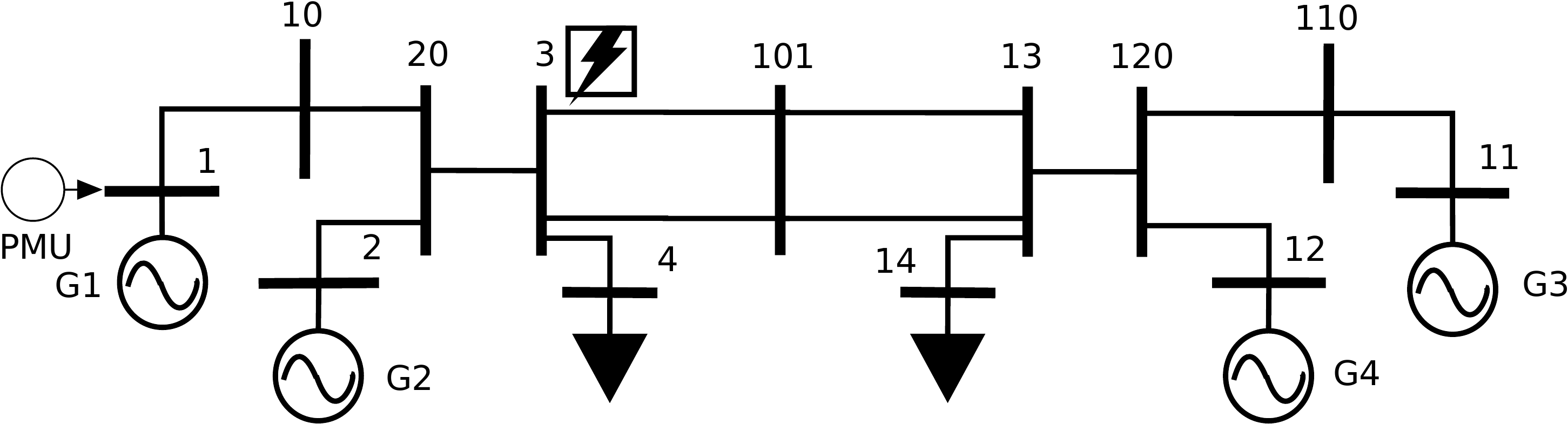}
\caption{Single line diagram of the test system.}
\label{fig:test_system}
\end{figure}

As it was mentioned before, the PST toolbox is used to perform all the simulations. The classical two-area and four machine system shown in Fig. \ref{fig:test_system} is the system to be considered. From the PST output, the measurements of a PMU located at the bus number 1 are generated. These measurements include the voltage and current phasors, their time derivatives, the frequency, and the ROCOF. All of them are computed considering a reporting rate of $f_{r}=60$ fps and an additive white Gaussian noise (AWGN). This AWGN condition can be achieved using a preprocessing of the data, as is shown in \cite{vanfretti2015}, so that this assumption is not as strong as it seems. Then, the adjustment of the noise variance is made. It guarantees the following values for the standard metrics:
\begin{equation}  \label{eq:metrics}
\mathrm{TVE}=1\%, \,\,\mathrm{FE{}}=5\mathrm{\,mHz},\,\, \mathrm{RFE}=0.1\,\mathrm{Hz/s}.
\end{equation}
In this manner, the measurement covariance matrix is defined:\footnote{Note that $\mathrm{diag}\{\mathbf{d}\}$ is a diagonal matrix with diagonal elements from $\mathbf{d}$ and the operator $\mathrm{cov}(\mathbf{d})$ is the covariance matrix of the random vector $\mathbf{d}$.}
\begin{equation}
 R_k=\mathrm{diag}\left\{\left[
\sigma_{V_{re}}^{2} \,\,  \sigma_{V_{im}}^{2} \,\, \sigma_{f}^{2} \,\, \sigma_{\alpha}^{2}\right]\right\}. \label{eq:meas_matrix}
\end{equation}

The voltage noise is complex and circularly-symmetric, i.e, $\sigma_{V_{re}}=\sigma_{V_{im}}=|\overrightarrow{V}|\,\mathrm{TVE}/(3\sqrt{2})$, and assuming \mbox{$|\overrightarrow{V}| \approx 1$ p.u}. The values of $\sigma_{f}$ and $\sigma_{\mathrm{\alpha}}$ were chosen from Monte Carlo simulations under realistic conditions, and they result to be $\sigma_{f}=\sigma_{\mathrm{\alpha}}=\sqrt{10^{-5}}$.

Samples are interpolated to reduce the effect of the system nonlinearities in the filtering stage. A linear interpolation is chosen with the following interpolation factor \mbox{$k_{int}=16$}. So, the time between samples results in: \mbox{$\Delta t=1.042 \times 10^{-3}$s}. The results presented above were obtained by proposing a realistic initial condition:
\begin{equation}
\mathbf{x}_0=\left[1.1\, \theta_0\,\,f_0\,\,\alpha_0\,\,1\,\,1\,\,1\,\,1\,\,1\,\,1\,\,0\,\,0.8\,H \,\, 0.6 \,K_A \right]^{T},
\end{equation} 
where $\theta_0$, $f_0$ and $\alpha_0$ are the phasor phase, the frequency and the ROCOF measurements 33ms after the fault is cleared. At this time, the large nonlinearities are reduced and the calibration process starts. Then, with this configuration, the following state covariance matrix and process noise covariance are defined as: 
\begin{equation}
Q_k=10^{-10}\,\Delta t\,\,I,\,\,P_{0}=\mathrm{cov}(\mathbf{p}),
\end{equation} 
\begin{equation}
\mathbf{p}\sim\mathcal{U}\left(\mathbf{x}_{\mathrm{0}\,\mathrm{ref}}-\Delta{\mathbf{x_\mathrm{0}},\mathbf{x}_{\mathrm{0}\,\mathrm{ref}}+\Delta{\mathbf{x_\mathrm{0}}}}\right),
\end{equation}
where  $\mathbf{x}_{\mathrm{0\,ref}}$ is the reference value of the state vector at the start time, $\Delta{\mathbf{x_\mathrm{0}}}=|\mathbf{x_{\mathrm{0}\,\mathrm{ref}}}-\mathbf{x_{\mathrm{0}}}|$ where $|\cdot|$ is the componentwise absolute value function, $\mathbf{p}$ is a random vector with uniform distribution and independent components, and $I$ is the identity matrix of $12\times12$. Finally, the selected parameters of the UKF are $\gamma=10^{-3}, \beta=2, \kappa=0$.

\begin{figure}[!t]
\centering
\includegraphics[width = 1\columnwidth]{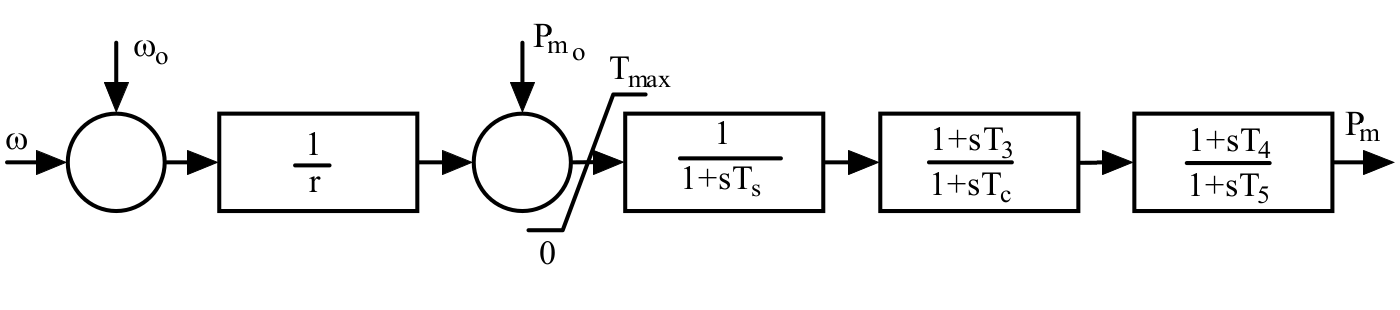}
\caption{Simple turbine governor model.}
\label{fig:tg_model1}
\end{figure}

\subsection{Scenario A}

\begin{figure}[!t]
\includegraphics[width = 0.95\columnwidth]{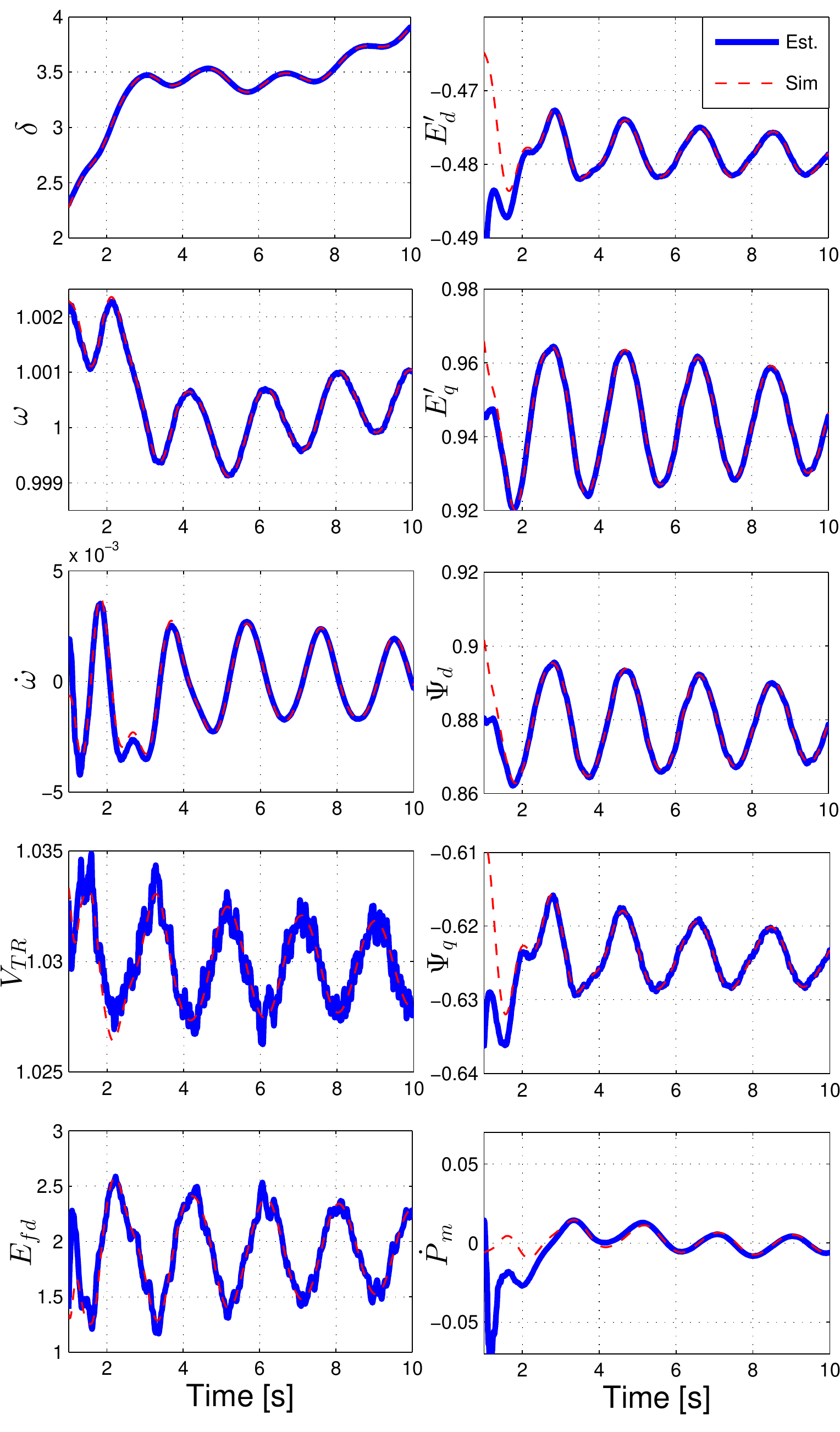}
\caption{Scenario A: Estimated dynamic states as a function of time.}
\label{fig:dynamic_states}
\end{figure}

\begin{figure}[t]
\centering
\includegraphics[width =.9\columnwidth]{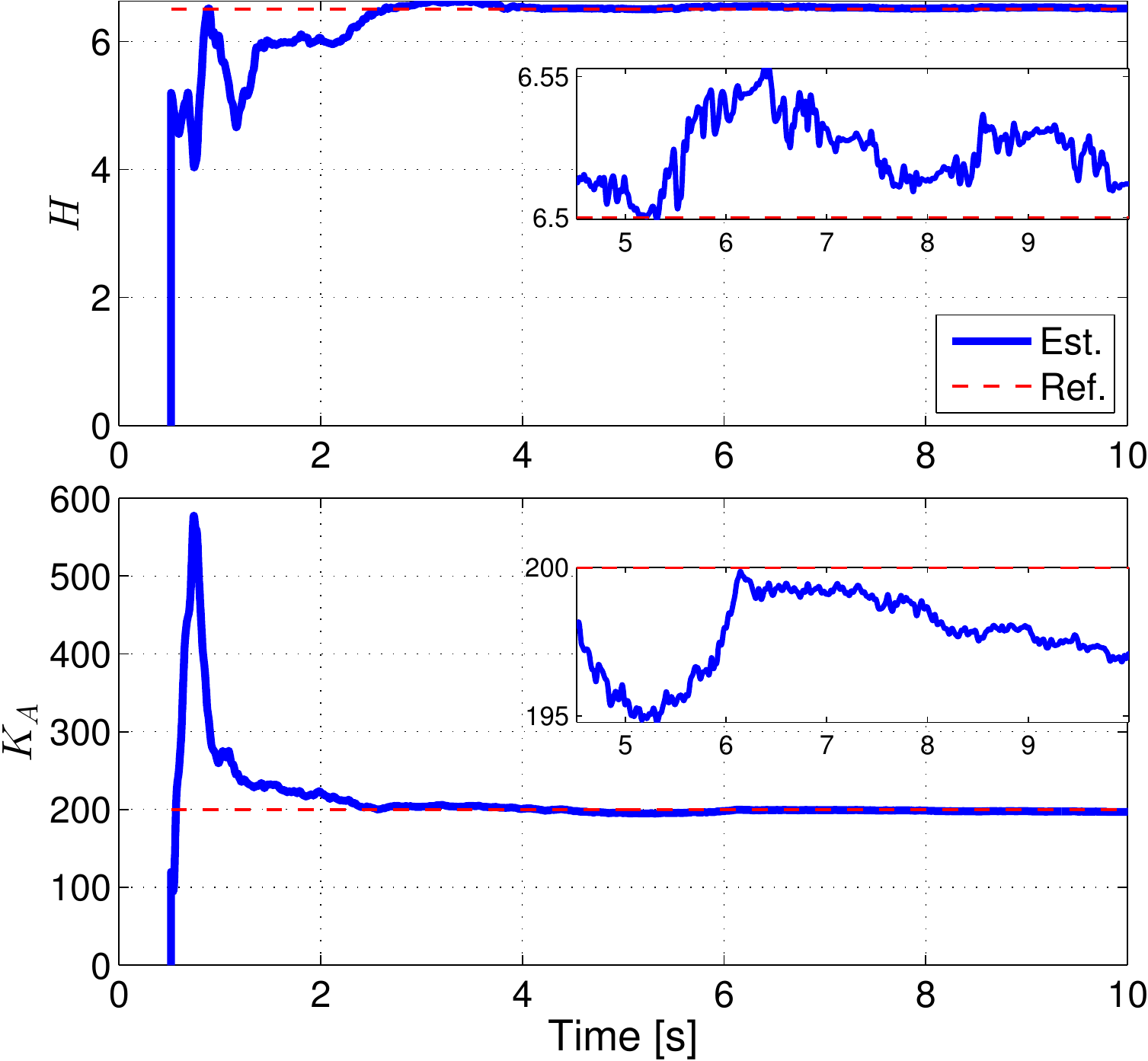}
\caption{Scenario A: Calibrated parameters as a function of time. A zoom of the state variable  after convergence is included.}
\label{fig:inertia_constant1}
\end{figure}

A three-phase line to ground fault is applied in the bus number 3 at \mbox{$t=0.1$}s. After another 100ms the fault is cleared and the system starts an oscillatory process. A sub-transient generator model for all generators is assumed (concretely the parameters used in the \textit{d2asbeg.m} PST example file). Using the model described in Section \ref{sec:sub_gen}, all the equations are matched with the simulator, except for the TG equations. For this subsystem PST considers a more sophisticated model that can be observed in Fig. \ref{fig:tg_model1}. As it can be seen, the transfer function of this subsystem implies more than one single pole. Several other parameters are included. From the transfer function, the associated cutoff frequency is calculated and the value of $T_{ef}$ is determined ($T_{ef}=2.4s$). Using the model presented in Section \ref{sec:improvement} the UKF is implemented, and the results are displayed in Fig. \ref{fig:dynamic_states} and \ref{fig:inertia_constant1}. It can be observed, that despite the noise added to the measurements and the poor initialization of the system, the dynamic state variables and the generator parameters are tracked with a good degree of accuracy.

\begin{table}[t]
\centering
\caption{Difference in \% with respect to the reference parameter.}
\label{tab:H and K parameters}
\begin{tabular}{|c|c|c|c|c|}
\hline
      & G1  & G2 & G3 & G4  \\ \hline
$H_0$   & -20 & 15 & 50 & -40 \\ \hline
$K_{A\,0}$ & -40 & 50 & 15 & -20 \\ \hline
\end{tabular}
\end{table}

\begin{table}[]
\centering
\caption{Scenario B: Mean and standard deviation of the estimated parameters ($H=6.5$,$K_A=200$), using the conventional model.}
\label{scenario_b1}
\begin{tabular}{|c|c|c|c|c|}
\hline
      & G1     & G2     & G3     & G4     \\ \hline
$\bar{H}$   & 7.3939 & 8.0854 & 1.8318 & 7.0461 \\ \hline
$s_H$ & 0.24177	& 0.34196	&11.312	&0.1473 \\ \hline
$\bar{K}_A$ & 182.29 & 188.5 & 198.67 & 190.28 \\ \hline
$s_{K_A}$ &3.8452	&2.1438	&26.806	&2.5905 \\ \hline
\end{tabular}
\end{table}

\subsection{Scenario B}
In order to evaluate the performance of a multiple generator calibration procedure, we propose to repeat the previous simulation for each one of the generator units (G1-G4), with the same calibration parameters. As in the previous case, the initialization of the dynamic variables is maintained, but the selection of the initial uncalibrated parameters is carried out arbitrarily, as detailed in Table \ref{tab:H and K parameters}. We perform a Monte Carlo simulation using the model of Section \ref{sec:improvement}, and we compare the results with other in the literature \cite{Zhou2015}. For that, the mean square error (MSE) is used as a metric.
\begin{equation}
\mathrm{MSE}(\hat{x}^j_k)=\frac{1}{M}\,\sum_{m=1}^{M} \,\left(\hat{x}^j_{k,m}-x^j_{k,\mathrm{ref}}\right)^2, \,\,j=1,\ldots,n,
\end{equation}
where $M=100$ is the number of Monte Carlo trials when the estimated $\mathrm{MSEs}$ did not change significantly, $\hat{x}^j_{k,m}$ is the $j$-th component of the state vector estimate at time $k$ and $m$-th realization, and $x^j_{k,\mathrm{ref}}$ is the true value of the state at the same time. Then, worst case of the $\mathrm{MSEs}$ after convergence is considered. To increase the dynamic range, \mbox{$\mathrm{MSE[dB]}=10 \,\,\log_{10}(\mathrm{MSE})$} is used to display the results in Tables \ref{scenario_b1} to \ref{scenario_b4}. The model of Section \ref{sec:sub_gen} can be contrasted against the one presented in Section \ref{sec:improvement}.  To make a fair comparison, a 2\% deviation of the value of the $P_{m,0}$ was added in the TG equation, which is an optimistic assumption, considering that the value in principle is unknown and it will always have an associated error. It is clear that for some realizations, the estimation based on the model G3 may be unstable. This will not be the case for the augmented model described in Section \ref{sec:improvement}.  

\begin{figure}[b]
\centering
\includegraphics[width = .95\columnwidth]{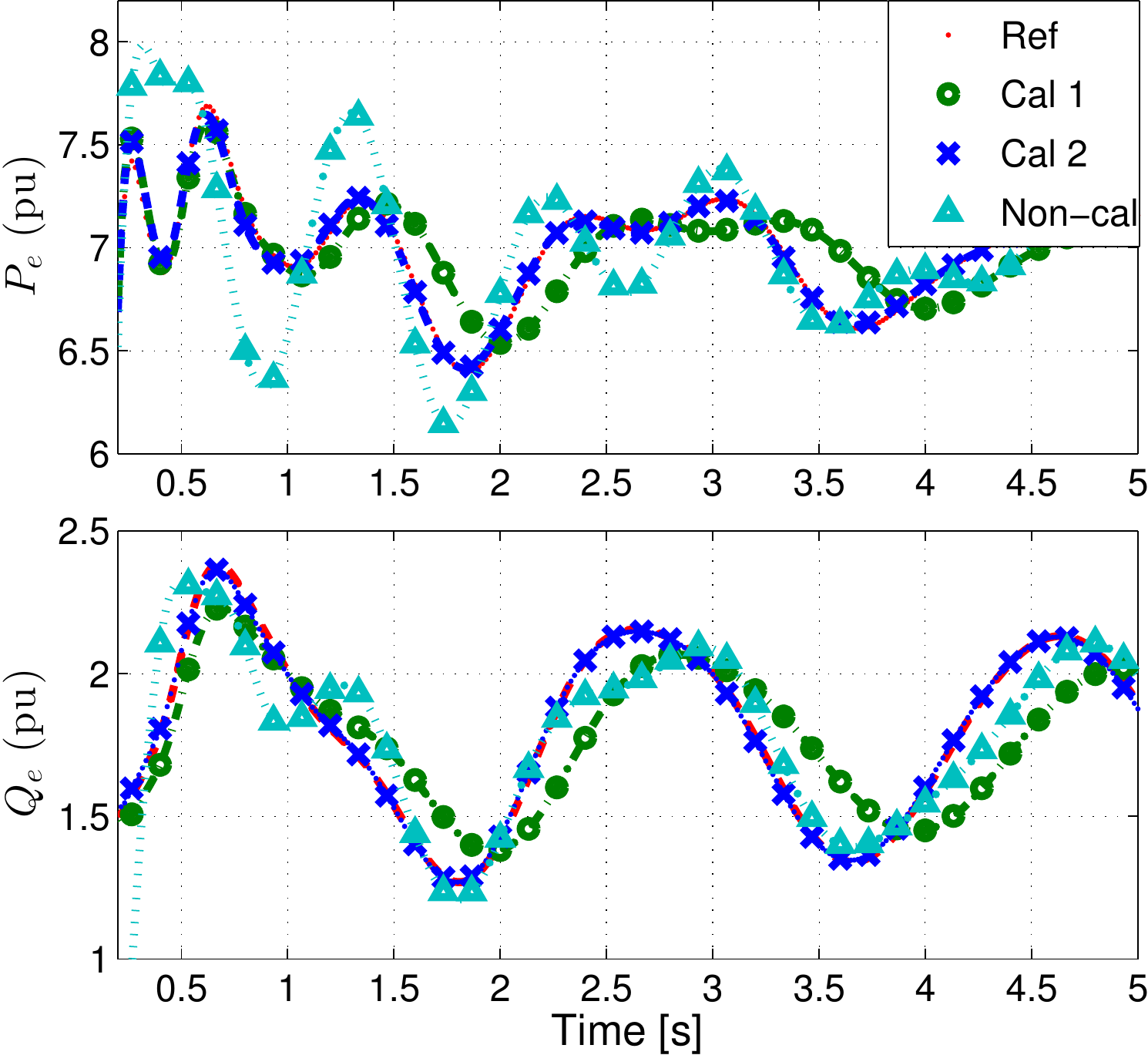}
\caption{Scenario B: Comparison among system outputs for different calibrations.}
\label{fig:inertia_constant2}
\end{figure}

It is relevant to analyze the output of the calibrated system and compare it with the one that is non-calibrated. In this context, the active ($P_e$) and reactive powers ($Q_e$) from G1 have been plotted in Fig. \ref{fig:inertia_constant2}. The label Cal 1 refers to estimates using the model described in Section \ref{sec:sub_gen}, while the label Cal 2 is reserved for the results from the augmented model. Again, the differences of the results are notorious.

\begin{table}[t]
\centering
\caption{Scenario B: MSE [$\mathrm{dB}$], using the conventional model.}
\label{scenario_b2}
\begin{tabular}{|c|c|c|c|c|}
\hline
State          & G1      & G2      & G3      & G4      \\ \hline
$\delta$      & -43.715 & -44.197 & 95.82   & -42.017 \\ \hline
$\omega$       & -74.615 & -74.94  & 15.221  & -73.68  \\ \hline
$E'_d$         & -66.845 & -69.245 & -23.005 & -69.665 \\ \hline
$E'_q$		   & -61.97  & -64.755 & -17.573 & -66.145\\ \hline
$\Psi_d$        &  -57.705 & -60.275 & -18.218 & -63.04 \\ \hline
$\Psi_q$        &  -58.215 & -59.47  & -20.905 & -58.175\\ \hline
$V_{TR}$       & -51.835 & -53.62  & -18.529 & -52.82 \\ \hline
$E_{fd}$ 	   &   -15.415 & -15.489 & -5.917  & -13.448 \\ \hline
$P_m$ 		   &  -36.189 & -36.155 & 52.265  & -36.272\\ \hline		
\end{tabular}
\end{table}

\begin{table}[t]
\centering
\caption{Scenario B: Mean and standard deviation of the estimated parameters ($H=6.5$,$K_A=200$), using the augmented model.}
\label{scenario_b3}
\begin{tabular}{|c|c|c|c|c|}
\hline
      & G1     & G2     & G3     & G4     \\ \hline
$\bar{H}$   & 6.6214   & 6.5513  & 6.1415  & 6.0334  \\ \hline
$s_H$ & 0.043687 & 0.12926 & 0.14719 & 0.16517 \\ \hline
$\bar{K}_A$ & 197.36   & 197.18  & 197.47  & 197.2   \\ \hline
$s_{K_A}$ & 2.2053   & 2.0048  & 2.1607  & 2.1406  \\ \hline
\end{tabular}
\end{table}

\begin{table}[b]
\centering
\caption{Scenario B: MSE [$\mathrm{dB}$], using the augmented model.}
\label{scenario_b4}
\begin{tabular}{|c|c|c|c|c|}
\hline
State          & G1      & G2      & G3      & G4      \\ \hline
$\delta$      & -57.88  & -59.975 & -61.935 & -61.36  \\ \hline
$\omega$       & -91.205 & -93.635 & -94.86  & -94.2   \\ \hline
$\dot{\omega}$ & -83.515 & -83.2   & -81.515 & -80.42  \\ \hline
$E'_d$         & -74.725 & -74.5   & -74.495 & -74.31  \\ \hline
$E'_q$         & -64.87  & -64.71  & -64.85  & -64.44  \\ \hline
$\Psi_d$        & -66.14  & -66.105 & -65.98  & -65.01  \\ \hline
$\Psi_q$        & -70.375 & -69.635 & -69.165 & -69.055 \\ \hline
$V_{TR}$       & -67.475 & -66.965 & -64.58  & -65.09 \\ \hline
$E_{fd}$ 	   &  -23.325 & -25.314 & -24.484 & -24.208\\ \hline
$\dot{P}_m$	   &  -72.885 & -74.265 & -66.815 & -71.165\\ \hline
\end{tabular}
\end{table}

\subsection{Scenario C}
\begin{figure}[t]
\centering
\includegraphics[width = 1\columnwidth]{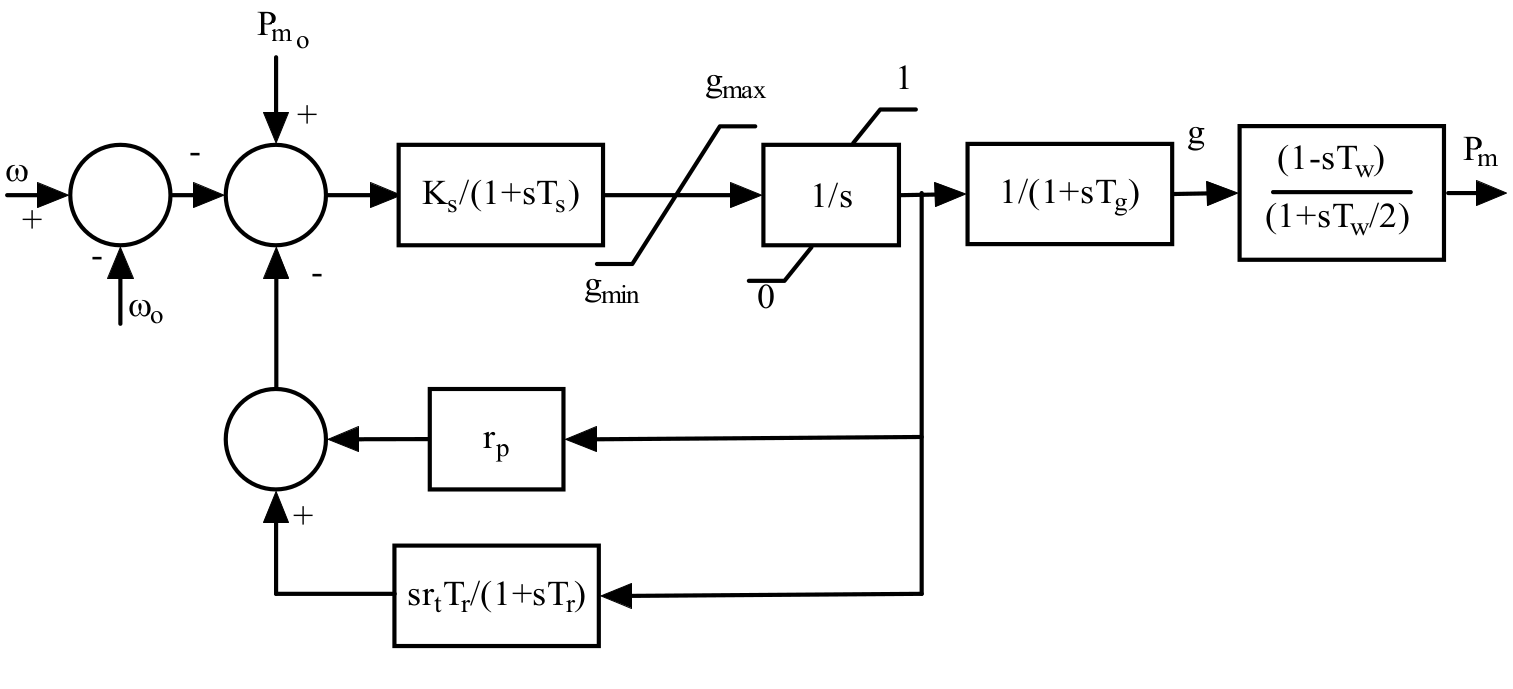}
\caption{Hydro turbine governor model.}
\label{fig:tg_model2}
\end{figure}

Finally, the TGs of all the generators are modified to test robustness against different models. Now, the Hydro-turbine model showed in Fig. \ref{fig:tg_model2} is selected (the parameters can be found in the PST example file \textit{d2asbegh.m}). It is clear that it presents a degree of complexity even more advanced than in the previous case. Without changing $T_{ef}$ or any initialization of the system, \cref{scenario_c1,scenario_c2} show the new results. Notice that performance is not significantly degraded.

\begin{table}[]
\centering
\caption{Scenario C: Mean and standard deviation of the estimated parameters ($H=6.5$,$K_A=200$), using the augmented model.}
\label{scenario_c2}
\begin{tabular}{|c|c|c|c|c|}
\hline
      & G1     & G2     & G3     & G4     \\ \hline
$\bar{H}$   & 6.6229  & 5.7946  & 6.2957 & 6.4214  \\ \hline
$s_H$ & 0.09272 & 0.37598 & 0.133  & 0.20374 \\ \hline
$\bar{K}_A$ & 202.73  & 203.1   & 203.08 & 202.84  \\ \hline
$s_{K_A}$ &2.2093  & 2.2296  & 2.2904 & 2.0347  \\  \hline
\end{tabular}
\end{table}

\begin{table}[]
\centering
\caption{Scenario C: MSE [$\mathrm{dB}$], using the augmented model.}
\label{scenario_c1}
\begin{tabular}{|c|c|c|c|c|}
\hline
State          & G1      & G2      & G3      & G4      \\ \hline
$\delta$      &-60.79  & -55.185 & -58.575 & -58     \\  \hline
$\omega$       &-96.505 & -85.665 & -91.665 & -90.74  \\  \hline
$\dot{\omega}$ &-86.69  & -71.6   & -79.15  & -77.575 \\  \hline
$E'_d$         &-75.37  & -75.38  & -74.305 & -75.235 \\  \hline
$E'_q$         &-65.02  & -66.17  & -64.73  & -65.97  \\  \hline
$\Psi_d$        &-66.16  & -68.335 & -66.385 & -66.875 \\  \hline
$\Psi_q$        &-69.925 & -68.85  & -68.57  & -69.185 \\  \hline
$V_{TR}$       &-66.35  & -65.99  & -64.165 & -65.315 \\  \hline
$E_{fd}$ 	   &-25.037 & -22.489 & -23.586 & -23.295 \\  \hline
$\dot{P}_m$	   &-64.515 & -64.295 & -59.86  & -60.495 \\  \hline
\end{tabular}
\end{table}

\section{Conclusions}
In this paper, we have introduced a new state variable model for a generator unit based on PMU data. For that, we have proposed to use knowledge of the voltage and current synchrophasors in conjunction with their time derivatives.

As it was mentioned in Section \ref{sec:improvement}, this approach takes advantage of the slow dynamics of the 
turbine governors. It is concluded that if the transition equations are based on the time derivative of the electric torque, the system is more robust and suitable for scenarios where the TG model is complex and only a coarse approximation of it is available. In turn, the TG model is simple and does not depend on a large number of parameters that are difficult to know a priori. In addition, the use of phasor derivatives allowed us to obtain a higher-order rotor dynamic model, which leads to an enhanced tracking of its state variables.

In Section \ref{sec:results}, the results have shown that it is feasible to perform parameter estimation and a dynamic tracking of the state variables simultaneously. In particular, scenario B shows that poor performance is obtained when using a conventional model. This poor performance produces an considerable difference between the actual output of the system and the simulated one after the calibration process. This is an important fact, since a bad simulation of the system can lead to poor network planning. Furthermore, scenario C has shown  decent results even though a complex hydro TG was used. 

Finally, we would like to emphasize once again that this approach can be expanded and used with other estimation techniques, especially with different Bayesian filters where the application is immediate.


%

\appendices




\section{Complementary equations}
\noindent
In this appendix, we list the complementary equations of the model defined by \eqref{eq:sub_tra1}-\eqref{eq:sub_tra4} and the expressions for additional model constants:
\begin{subequations}
\begin{align}
I_{d} & =I_{re}\sin\left(\delta\right)-I_{im}\cos\left(\delta\right), \label{eq:id}\\
I_{q} & =I_{im}\sin\left(\delta\right)+I_{re}\cos\left(\delta\right), \label{eq:iq}\\
T_{e} & =P_{e}\,+\,r_{A}\,\left(I_{d}^{2}+I_{q}^{2}\right), \label{eq:electric_torque}\\
E_{d} & =\Psi''_{q}-r_{A}I_{d}+x''_{d}I_{q}\,,\\
E_{q} & =\Psi''_{d}-r_{A}I_{q}-x''_{d}I_{d}\,,\\
V & =\sqrt{E_{d}^{2}+E_{q}^{2}}\,,\\
S &=k_{sat1} E'_q{}^2 + k_{sat2} E'_q + k_{sat3}.
\end{align}
\end{subequations}
\begin{eqnarray*}
k_1 =\frac{(x_q-x'_q)(x'_q-x''_q)}{(x'_q-x_{ls})^2}, &
k_2 =\frac{(x_q-x_{ls})(x''_q-x_{ls})}{(x'_q-x_{ls})},\\
k_3 =\frac{(x_d-x'_d)(x'_d-x''_d)}{(x'_d-x_{ls})^2}, &
k_4 =\frac{(x_d-x'_{d})(x''_d-x_{ls})}{(x'_d-x_{ls})}.
\end{eqnarray*}

\section{Nomenclature}
%
\begin{table}[!h]
\centering
\normalsize
\begin{tabular}{|c|c|}
\hline
\multicolumn{2}{|c|}{\textit{Variables and constants [p.u.]}}\tabularnewline \hline
$\delta$ &     Rotor angle.  \tabularnewline \hline
$\omega$ &     Angular velocity of the rotor.   \tabularnewline \hline
$\dot{\omega}$ &     Time derivative of rotor velocity.  \tabularnewline \hline
$f$ &     Instantaneous system frequency. \tabularnewline \hline
$\alpha$ &     Instantaneous system ROCOF.  \tabularnewline \hline
$\Psi_{d}/\Psi_{q}$ &   d/q axis subtransient voltage.  \tabularnewline \hline
 $E'_{d}/E'_{q}$ &     d/q axis transient voltage.  \tabularnewline \hline
$P_{e}/\,T_{e}$ &     Active electric power / torque.  \tabularnewline \hline
$P_{m}/P_{m,0}$ &   Instantaneous/Steady state mechanical power  \tabularnewline \hline
$E_{fd}$ &    Field voltage.  \tabularnewline \hline
$I_{d}/I_{q}$ &    d/q axis stator current.  \tabularnewline \hline
$V_{TR}$ &    Transducer output signal.  \tabularnewline \hline
$pss$ &     Power system stabilizer signal.  \tabularnewline \hline
$S$ &     q-axis saturation function.  \tabularnewline \hline
$\omega_{0}$ &     Nominal rotor speed (1 p.u.)  \tabularnewline \hline
$r_{A}$ &     Stator resistance  \tabularnewline \hline
$x_{d},x{}_{q}$ &     d/q axis synchronous reactance  \tabularnewline \hline
$x'_{d},x'_{q}$ &     d/q axis transient reactance  \tabularnewline \hline
$x''_{d},x''_{q}$ &    d/q axis sub-transient reactance  \tabularnewline \hline
$x_{ls}$ &     Stator leakage reactance  \tabularnewline \hline
$T'_{d},T'_{q}$ &     d/q axis transient open circuit time constant  \tabularnewline \hline
$T''_{d},T''_{q}$ &     d/q axis subt-tran. open circuit time constant  \tabularnewline \hline
$k_{sati}$ &     i-th core saturation factors  \tabularnewline \hline
$D$ &     Damping factor  \tabularnewline \hline
$H$ &     Inertia constant  \tabularnewline \hline
$K_{A}/\frac{1}{r}$ &     Exciter/Turbine governor gain   \tabularnewline \hline
$T_{R}$ &     Exciter time constant  \tabularnewline \hline
$T_{ef}$ &      Turbine governor effective time constant  \tabularnewline \hline
$V_{REF}$ &     Reference voltage of the excitation system  \tabularnewline \hline
\end{tabular}
\end{table}

\ifCLASSOPTIONcaptionsoff
  \newpage
\fi



\bibliographystyle{IEEEtran}
\bibliography{biblio}
\end{document}